\documentclass[amstex,pra,preprint]{revtex4}%
\usepackage{amsmath}
\usepackage{amsfonts}
\usepackage{amssymb}
\usepackage{graphicx}
\usepackage{xcolor}%
 
\begin{document}
\title{Proposal to observe paths superpositions in a double-slit setup}
\author{Q.\ Duprey}
\affiliation{ENSEA, 6 avenue du Ponceau, 95014 Cergy-Pontoise cedex, France}
\author{A. Matzkin}
\affiliation{Laboratoire de Physique Th\'eorique et Mod\'elisation, CNRS
Unit\'e \ 8089, CY Cergy Paris Universit\'e, 95302 Cergy-Pontoise
cedex, France}

\begin{abstract}
The interference pattern produced by a quantum particle in Young's double-slit
setup is attributed to the particle's wavefunction having gone through both
slits. In the path integral formulation, this interference involves a
superposition of paths, going through either slit, linking the source to the
detection point. We show how these paths superpositions can in principle be
observed by implementing a series of minimally-perturbing weak measurements
between the slits and the detection plane. We further propose a simplified
protocol in order to observe these ``weak trajectories'' with single photons.

\end{abstract}
\maketitle


\section{Introduction}

The Weak Measurement (WM) scheme \cite{origine} is a minimally-disturbing
process that enables one to gain information on the property of a quantum
system at an intermediate time $t$ as the system evolves from an initially
prepared state $|\psi(t_{i})\rangle$ to a state $|\chi(t_{f})\rangle$ obtained
as the result of a projective measurement at time $t_{f}$. More specifically,
the WM of an observable $\hat{O}$ is achieved by introducing a weak coupling
at an intermediate time of the evolution : a dynamical variable of the quantum
pointer is weakly coupled at time $t$ to the observable $\hat{O}$ of the
system (a quantum particle) such that the probability of transition between
the initial (``pre-selected'') state $|\psi(t_{i})\rangle$ and the detected
(``post-selected'') state $|\chi(t_{f})\rangle$ remains unchanged relative to
the no-coupling case. When the final measurement at $t=t_{f}$ projects the
system to state $|\chi(t_{f})\rangle$, the weak value of $\hat{O}$ can be read
by measuring the quantum pointer. The weak value $O_{w}$ is a complex quantity
given by
\begin{equation}
O^{w}(t)=\frac{\langle\chi(t_{f})|\hat{U}(t_{f},t)\hat{O}\hat{U}(t,t_{i}%
)|\psi(t_{i})\rangle}{\langle\chi(t_{f})|\hat{U}(t_{f},t_{i})|\psi
(t_{i})\rangle} \label{wvd}%
\end{equation}
where $\hat{U}(t^{\prime},t)$ denotes the evolution operator of the system
from $t$ to $t^{\prime}$. Typically, the momentum of the pointer is coupled to
$\hat{O}$, in which case it can be shown \cite{origine,sudarshan} that the
initial pointer state is shifted by a quantity proportional to
$\operatorname{Re}O^{w}$.

When $\hat{O}$ is given by the position operator $\hat{X}$, or by the
projection to a particular spatial position $\Pi_{x}\equiv\left\vert
x\right\rangle \left\langle x\right\vert $ , the weak value captures
information about the position of the system as it evolves between the pre-
and post-selected states. By inserting in different spatial regions a series
of weakly coupled pointers that can be turned on and off at the desired time,
it is possible to define ``weak trajectories'' \cite{matzkin}.\ Such
trajectories are in principle observable signatures of the space-time
evolution of the system. In simple cases, e. g. with narrow coherent states
\cite{tanaka,matzkin} or with point-like pre and post-selected states
\cite{mori1} and free propagation, the system is seen to follow a classical
trajectory, or a superposition of such trajectories.

This is reminiscent of Feynman's path integral approach \cite{feynman}, in
which the evolution operator can be written (for free propagation, or when the
potential is linear or quadratic in $x$) as a sum over classical paths
\cite{schulman}. In more complex situations
\cite{matzkinPR,mori2,narducci,hiley,georgiev,katz}, several Feynman paths
compatible with pre- and post-selection can interfere at the position of a
weakly coupled probe, and may even result in a vanishing weak value (e.g.,
\cite{papiernull}). In a generic situation, Feynman paths interfere as the
system evolves, so that reconstructing a weak trajectory from observed
pointers is expected to be a difficult task.

The double-slit interferometer is arguably the simplest system displaying
interference of Feynman paths: a point on the detection screen involves
interferences between two paths, one coming from each slit. Aspects of weak
trajectories in a double-slit setup were recently investigated
\cite{mori1,mori2,narducci} from an ideal point of view (point-like pre and
post-selected states, absence of weakly coupled pointers). In this work, we
will propose a protocol to observe weak trajectories in view of a possible
future experimental implementation with single photons.\ Indeed, the weak
coupling of a quantum particle with an array of independent quantum pointers
looks hardly feasible with current technologies. However, weak measurements
have been successfully implemented on single photons by employing couplings
with their polarization degree of freedom instead of genuine quantum pointers;
this was done in particular in a double slit setup that measured the velocity
field of single photons \cite{kocsis}, related to the weak value of the momentum.

Similarly, we will present here a protocol in order to detect with weak
measurements the paths superposition contained in the Feynman propagator in a
double-slit interferometer. We will first briefly review the free Feynman
propagator and the associated paths for a double-slit setup (Sec.
\ref{sec-2}). We will then introduce in Sec. \ref{secideal} weak trajectories
in an ideal scenario in which the wavefunction at each slit can be controlled,
and then weakly interacts with a series of probes. This ideal scenario is
useful in order to illustrate in a simple manner the superposition of weak
trajectories. In a genuine double-slit setup however, the wavefront is carried
by a sum over many interfering paths; how the corresponding weak trajectories
could in principle be recovered by relying on weak measurements is studied in
Sec. \ref{scheme}. Such a scheme is probably not realizable with current
technologies. We therefore introduce (Sec. \ref{photonic}) a protocol having
in mind photonic experiments for which it should be possible to observe weak
trajectories with currently available tools. This protocol is based on the
interaction between a single photon and a minimal number of birefringent
crystals placed between the slits and the detection screen. We discuss our
results and conclude in Sec. \ref{sec-conc}.

\begin{figure}[ptb]
\centering \includegraphics[width=8.2cm]{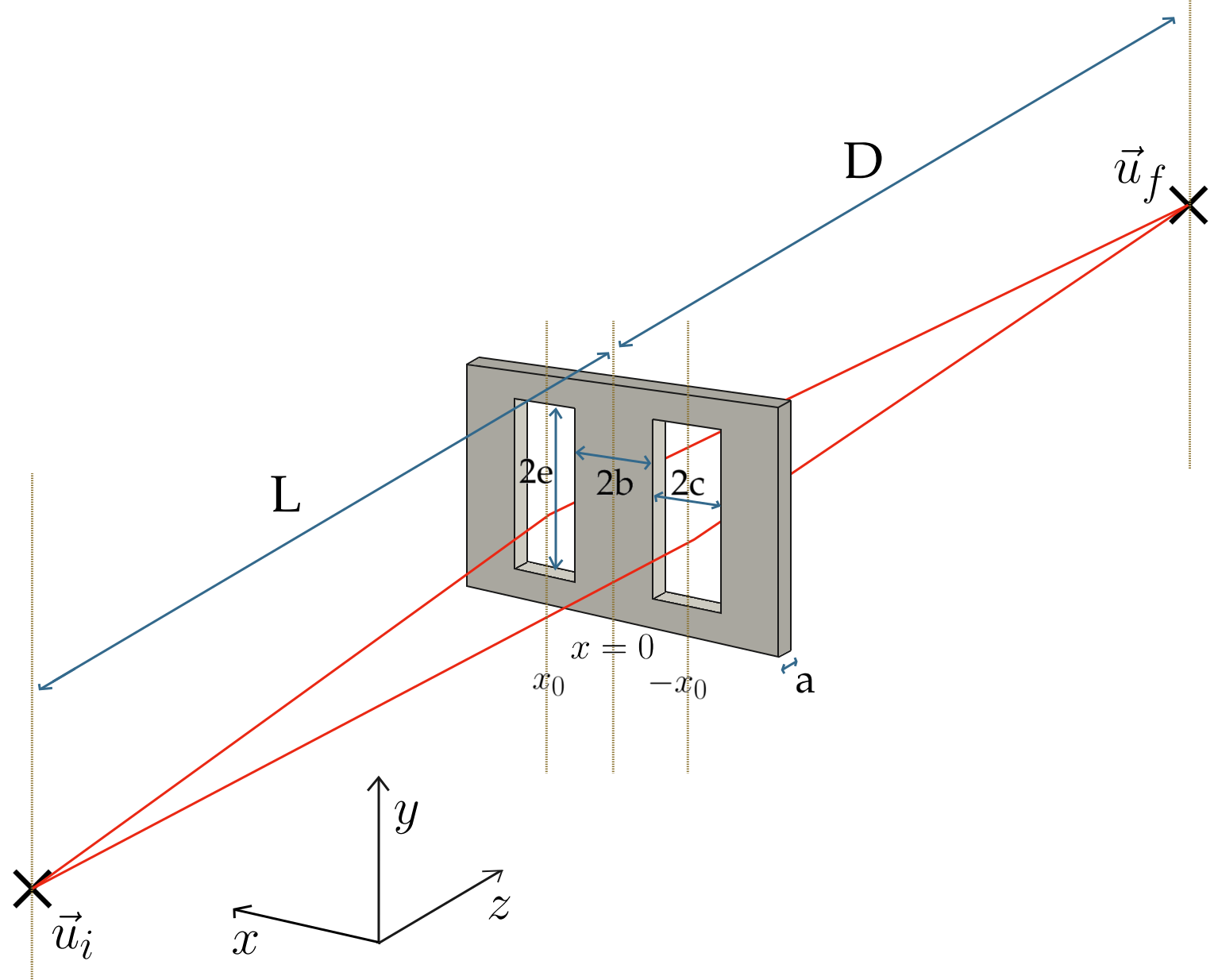} \caption{Young's double-slit
setup, introducing the notation employed in the text.}%
\label{fentes}%
\end{figure}

\section{Feynman paths and propagator in a double-slit interferometer
\label{sec-2}}

\label{propagatorsection} Let us consider Young's double slit setup
represented in Fig. \ref{fentes}. A particle is emitted at time $t_{i}$ by a
point source located at $\vec{u}_{i}$ and detected at time $t_{f}$ at point
$\vec{u}_{f}$ of a screen. The slits of width $2c$ and of thickness $a$ are
centered on $\pm x_{0}$ where $2x_{0}=2b+c$. We base our treatment on the
approximations that are usually made \cite{kobe,beau,beau2} in this problem, namely:

\begin{itemize}
\item sufficiently long slits in the $y$ direction so that diffraction effects
along that direction can be neglected (hence only the propagator in the $xz$
plane becomes relevant);

\item factor the $xz$ propagation into a plane-wave propagation along $z$ and
a propagator along $x.$ This approximation, sometimes known as the truncation
approximation \cite{troncation}, is justified in the Fraunhofer regime ($L\gg
c$) and when the particle momentum is essentially along $z$ $(\left\vert
p_{z}\right\vert \gg\left\vert p_{x}\right\vert )$;

\item the thickness $a$ is deemed negligible, and the contribution of the loop
trajectories \cite{yakubi,sinha} (going several times through the slits) are
not taken into account.
\end{itemize}

Under these assumptions, the wavefunction of the problem is of the form
$\Psi(x,z,t)=\psi(x,t)e^{ip_{z}z}e^{-i\frac{p_{z}^{2}}{2m}(t-t_{i})/\hbar}%
$\ and the relevant interference pattern is the one along the $x$%
-direction.\ $\psi(x,t)$ is computed from the initial $\psi(x,t_{i})$ through
the propagator,
\begin{equation}
\psi(x^{\prime},t^{\prime})=\int_{-\infty}^{\infty}dx\hspace{0.05cm}%
K(x^{\prime},t^{\prime}|x,t_{i})\psi(x,t_{i}),
\end{equation}
where
\begin{equation}
K(x^{\prime},t^{\prime}|x,t)\equiv\left\langle x^{\prime}\right\vert
e^{-i\frac{\hat{p}^{2}}{2m}(t^{\prime}-t)}|x\rangle=\left(  \frac{m}%
{2i\pi\hbar(t^{\prime}-t)}\right)  ^{\frac{1}{2}}\hspace{0.05cm}%
e^{\frac{im(x^{\prime}-x)^{2}}{2\hbar(t^{\prime}-t)}} \label{amplitudelibre}%
\end{equation}
is the propagator for the free Hamiltonian $\hat{H}=\hat{p_{x}}^{2}/2m$. Eq.
(\ref{amplitudelibre}) is obtained from the path integral by \textquotedblleft
slicing\textquotedblright\ the evolution operator, and as is well-known
\cite{schulman} the result is that all the paths \ interfere destructively
except the classical trajectory linking the point $(x,t)$ to $(x^{\prime
},t^{\prime})$. The phase in the exponent is the classical action
$s(x^{\prime},t^{\prime})$ of that trajectory, while the amplitude of the
propagator depends on the classical density $\partial^{2}s/\partial x\partial
x^{\prime}$ of that trajectory.

We consider the slits as openings in a potential barrier. The functions
$\varphi_{1}(x)$ and $\varphi_{2}(x)$ represent the shape of the slits 1 and 2
centered on $x_{0}$ and $-x_{0}$ respectively. Assuming a point-like source,
i.e. $\Psi(x_{i},z_{i},t_{i})=\delta(x-x_{i})\delta(z-z_{i})$, the propagator
from the source to a point $\left(  x_{f},t_{f}\right)  $ beyond the slits
take the form
\begin{align}
K(x_{f},t_{f}|x_{i},t_{i})  &  =\int_{-\infty}^{\infty}dx\hspace
{0.05cm}K(x_{f},t_{f}|x,\tau)(\varphi_{1}(x)+\varphi_{2}(x))K(x,\tau
|x_{i},t_{i})\\
&  =K_{1}(x_{f},t_{f}|x_{i},t_{i})+K_{2}(x_{f},t_{f}|x_{i},t_{i}) \label{prop}%
\end{align}
where we have defined $K_{j}(x_{f},t_{f}|x_{i},t_{i})=\int_{-\infty}^{\infty
}dx\hspace{0.05cm}K(x_{f},t_{f}|x,\tau)\varphi_{j}(x)K(x,t_{f}|x_{i},t_{i})$
to be the propagator associated with a passage through slit $j$ at time $\tau$.

For mathematical simplicity, we will take $\varphi_{j}(x)$ to be Gaussian
functions $\varphi_{1}(x)=e^{-\frac{(x-x_{0})^{2}}{2c^{2}}}$ and $\varphi
_{2}(x)=e^{-\frac{(x+x_{0})^{2}}{2c^{2}}}$ \cite{kobe}, rather than
rectangular windows \cite{beau}. Taking $x_{i}=0$ and $t_{i}=0$, $K_{j}%
(x_{f},t_{f}|x_{i},t_{i})$ becomes \cite{kobe}
\begin{equation}%
\begin{split}
K_{j}(x_{f},t_{f}|0,0)=  &  \bigg\lbrace\frac{m}{2i\pi\hbar(t_{f}+i\alpha
\tau(t_{f}-\tau))}\bigg\rbrace^{\frac{1}{2}}\\
&  \text{exp}\bigg\lbrace-\frac{1}{2(\Delta x)^{2}}(1-i\eta)(x_{f} -
\delta_{j} x_{0}\frac{t_{f}}{\tau})^{2}+i\beta\frac{(x_{f}- \delta_{j}
x_{0})^{2}}{t_{f}-\tau}+i\beta\frac{x_{0}^{2}}{\tau}\bigg\rbrace\label{kx2}%
\end{split}
\end{equation}
where
\begin{equation}%
\begin{split}
(\Delta x)^{2}=\bigg(c\frac{t_{f}}{\tau}\bigg)^{2}+\bigg(\frac{\hbar
(t_{f}-\tau)}{mc}\bigg)^{2}\quad\beta=\frac{m}{2\hbar}\quad\alpha=\frac{\hbar
}{mc^{2}}\\
\eta=\frac{ct_{f}/\tau}{\hbar(t_{f}-\tau)/mc} \quad\delta_{j=1}=+1 \quad
\delta_{j=2}=-1.
\end{split}
\end{equation}
The resulting probability at $\left(  x_{f},t_{f}\right)  $ is readily
obtained from Eqs. (\ref{prop}) and (\ref{kx2}). In the Fraunhofer
approximation, $\Delta x$ is greater than $\langle x_{f}\rangle$ and the
probability at the screen $\mathcal{P}(x_{f},t_{f})$ can be approximated to
the compact form \cite{kobe}
\begin{equation}
\mathcal{P}(x_{f},t_{f})=\cos^{2}\bigg(\frac{2p_{z}x_{0}x_{f}}{\hbar
D}\bigg)e^{-\frac{x_{f}^{2}}{(\Delta x)^{2}}}%
\end{equation}
where $p_{z}=mv_{z}=mD/(t_{f}-\tau)$.

This interference is due to the superposition of the two type of trajectories
associated with Eq. (\ref{prop}): those that go from $x_{i}$ to $x_{f}$
through slit 1, and those that go through slit 2. In principle, there is an
infinite number of classical trajectories associated with each propagator
$K_{j}(x_{f},t_{f}|x_{i},t_{i})$, each distinguished by a different
intermediate point on the slit plane.\ Note however that Eq. (\ref{kx2})
indicates that each $K_{j}$ is essentially a Gaussian wavefunction that can be
characterized by the classical trajectory followed by the maximum of the
Gaussian, $x_{j}(t)=\delta_{j}x_{0}+\frac{p_{x}^{j}}{m}(t-\tau).$

\section{An ideal case: controlling pre-selection at the slits
\label{secideal}}

\subsection{An array of weakly coupled pointers}

We will see here how the superposition of paths going through each slit can be
obtained in an ideal double-slit experiment in which a set of quantum
pointers, disposed on a grid, may interact -- weakly and unitarily -- with the
system (say a single electron) originating from the source, and detected on
the screen. We will assume each probe is coupled to the spatial projection
operator $\hat{\Pi}_{a}\equiv\left\vert \mathbf{r}_{a}\right\rangle
\left\langle \mathbf{r}_{a}\right\vert $ of the system, where $\mathbf{r}%
_{a}=(x_{a},z_{a})$.\ The pre and post-selected states, to be specified later,
will be denoted by $\left\vert \psi(t_{i})\right\rangle $ and $\left\vert
\chi(t_{f})\right\rangle $ respectively. Applying the weak value definition
(\ref{wvd}) to $\hat{\Pi}_{a}$ for an interaction at time $t$ gives%

\begin{align}
\left(  \Pi_{a}\right)  ^{w}(t)  &  =\frac{\langle\chi(t_{f})|\hat{U}%
(t_{f},t)\hat{\Pi}_{a}\hat{U}(t,t_{i})|\psi(t_{i})\rangle}{\langle\chi
_{f}(t_{f})|\hat{U}(t_{f},t_{i})|\psi(t_{i})\rangle}\label{wv1}\\
&  =\frac{\chi^{\ast}(x_{a},t)\psi(x_{a},t)}{\int dx\chi^{\ast}(x,t)\psi
(x,t)}=\left(  \Pi_{x_{a}}\right)  ^{w}(t) \label{wv2}%
\end{align}
where $\hat{U}(t,t_{i})$ is the free evolution operator. Note that given the
assumptions introduced in Sec. \ref{sec-2}, the evolution along the $z$ axis
is trivial, yielding phase factors that cancel out, so that only the weak
value $\left(  \Pi_{x_{a}}\right)  ^{w}$ along $x$ is relevant. We will not
review here the properties of weak values (see e.g. \cite{matzkinFP} for a
recent review), though it should be recalled that $\left(  \Pi_{a}\right)
^{w}$ is complex-valued and only the real part (inducing a shift on a pointer)
is related to the property of the system observable $\hat{\Pi}_{a}$ coupled to
the pointer.

\begin{figure}[h]
\centering \includegraphics[width=8.4cm]{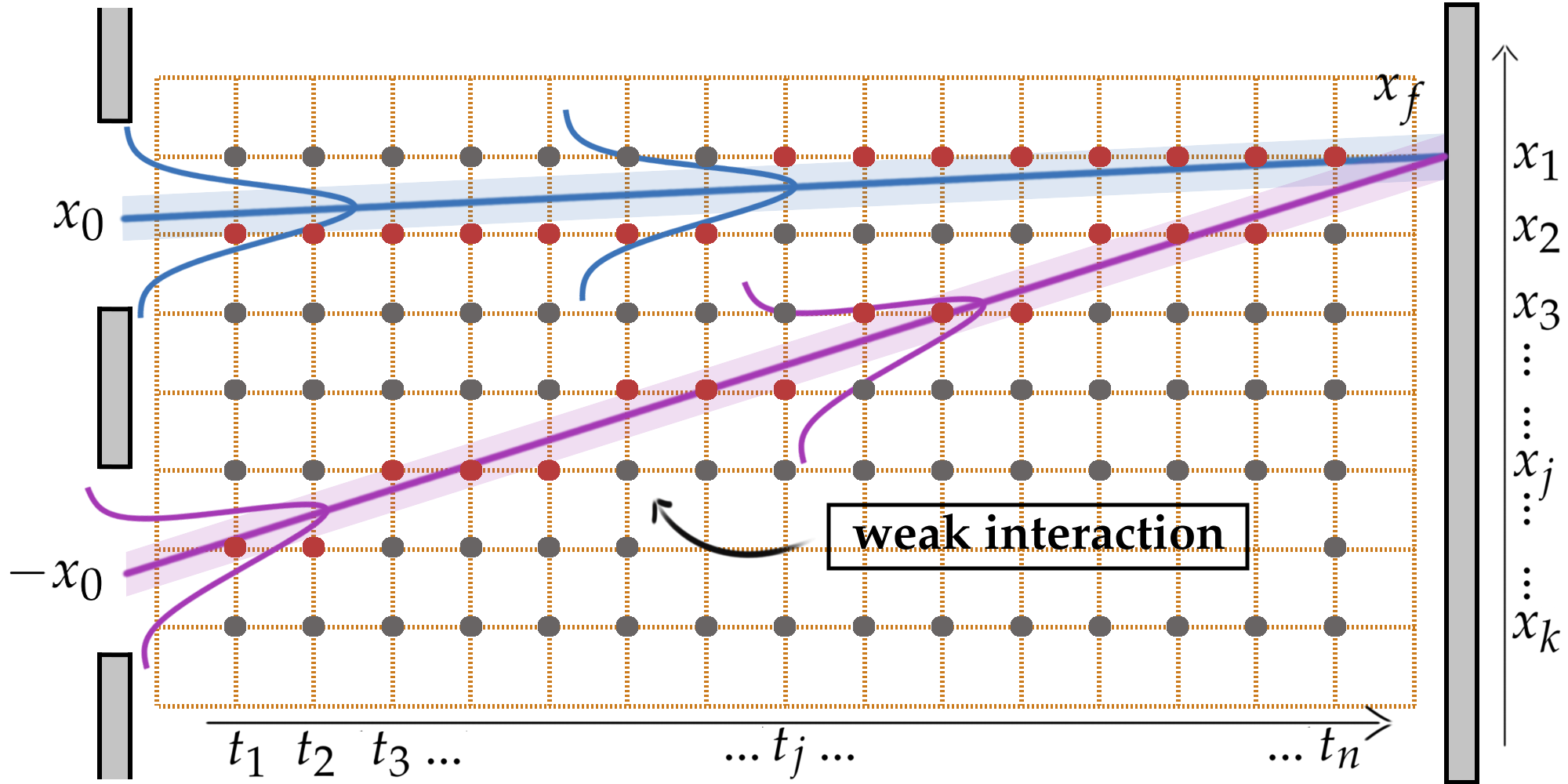}\caption{Ideal two-slits setup
with a grid of weakly coupled probes. A particle is initially in a
superposition of coherent states at the slits. The coherent states are
directed toward a point $x_{f}$ on the screen. Each dot represents a probe
that couples to the particle through a weak unitary interaction. Upon
post-selection, some probes see their quantum state shifted as a
result of the weak interaction. Each set $\{t_{k},\mathbf{r}_{k}\}$ of these
shifted probes defines a ``weak trajectory''. With an appropriate choice of
post-selection, the weak trajectories shown here joining slit 1 to $x_{f}$ and
slit 2 to $x_{f}$ can be detected jointly (see text for details).}%
\label{idealpaths}%
\end{figure}

The probes are organized on a grid as schematically shown in Fig.
\ref{idealpaths}. Let us label each probe by $\mathcal{M}_{\text{ab}}$, where
a stands for the position $(x_{a},z_{a})$ and b for the time of interaction
$t_{b}$. Let $|\xi_{a,b}\rangle$ denote the quantum state of the probe
$\mathcal{M}_{\text{ab}}$. We will assume the interaction Hamiltonian between
the system and each probe to be of the form
\begin{equation}
\hat{H}_{\text{int}}^{a,b}=g(t_{b})\hat{\Pi}_{a}\hat{P}_{a} \label{ham}%
\end{equation}
where $g(t_{b})$ is the coupling strength (a smooth function peaked at $t_{b}
$), and where $\hat{P}_{a}$ is a dynamical variable of probe $\mathcal{M}%
_{\text{ab}}$ (typically the momentum, but see below in Sec.\ \ref{photonic}
for a different choice). Setting $\int_{t_{b}-T/2}^{t_{b}+T/2}g(t^{\prime
})dt^{\prime}/\hbar\equiv\gamma_{b}$ where $T$ is the duration of the weak
coupling, the evolution operator due to the weak coupling between the system
and a probe becomes
\begin{equation}
\exp\left(  -i\gamma_{b}\hat{\Pi}_{a}.\hat{P}_{a}\right)  \simeq1-i\gamma
_{b}\hat{\Pi}_{a}.\hat{P}_{a}. \label{deva}%
\end{equation}
The $b$ subscript in $\gamma_{b}$ is a label added in order to distinguish the
interaction times wherever necessary (this has no bearing on the numerical
values of all the $\gamma_{b}$'s that will taken to be the same assuming
identical probes). With this notation, taking for convenience $t=0$ when the
system wavefunction is at the slits, the overall quantum state, given
initially by $\left\vert \Psi(t=0)\right\rangle =|\psi(t=0)\rangle
\bigotimes_{a}|\xi_{a,b}(t=0)\rangle$, becomes
\begin{equation}
|\Psi(t_{f}\rangle=\prod_{b=1}^{n}\left[  \hat{U}(t_{b},t_{b-1})\prod
_{a=1}^{k}e^{-i\gamma_{b}\hat{\Pi}_{a}.\hat{P}_{a}}\right]  \left\vert
\Psi(t=0)\right\rangle ,
\end{equation}
with $t_{0}=0$ and $t_{n}=t_{f}.$ Applying Eq. (\ref{deva}) and keeping only
terms to first order in the couplings $\gamma_{b}$, and finally post-selecting
to the state $\left\vert \chi_{f}\right\rangle $ leaves the quantum state of
each probe $\mathcal{M}_{\text{ab}}$ shifted in proportion to the
corresponding weak value,%
\begin{equation}
\left\vert \xi_{a,b}(t_{f})\right\rangle =\exp\left(  -i\gamma_{b}\left(
\Pi_{a}\right)  ^{w}\hat{P}_{a}\right)  \left\vert \xi_{a,b}(t=0)\right\rangle
. \label{idiprobe}%
\end{equation}
In principle, the quantum state of each probe can be read out after
post-selection. From the knowledge of the position $\mathbf{r}_{a}$ of each
probe and the control of the interaction time $t_{b},$ one can define ``weak
trajectories'' by linking the probes for which $\operatorname{Re}\left(
\Pi_{a}\right)  ^{w}\neq0 $.

\subsection{\bigskip Controlling pre-selection at the slits}

\label{prepostsection} \noindent We have seen in Sec. \ref{sec-2} that a point
source located behind Gaussian slits results in a Gaussian propagating from
each slit. In this Section we will introduce weak trajectories for slits in a
slightly more general situation in which we assume the Gaussian wavefunctions
emanating from the slits can be controlled.

To this end, consider the pre-selected state to be given by
\begin{equation}
\psi(x,t)=\frac{1}{\sqrt{2}}\bigg(\psi_{p_{i}^{1}}^{1}(x,t)+\psi_{p_{i}^{2}%
}^{2}(x,t)\bigg) \label{initial}%
\end{equation}
where the Gaussian%
\begin{equation}
\psi_{{}}^{j}(x,0)=\frac{1}{(2\pi d^{2})^{\frac{1}{4}}}e^{-i\frac
{(x-\delta_{j}x_{0})^{2}}{4d^{2}}+i\frac{p_{i}^{j}(x-\delta_{j}x_{0})}{\hbar}}
\label{gaussiennesslit}%
\end{equation}
is the wavefunction for slit $j$ (as above $\delta_{j}=+1$ for $j=1$ and
$\delta_{j}=-1$ for $j=2$). $p_{i}^{j}$ denotes the initial ($t=0$) mean
momentum along $x$ of the slit $j$ wavefunction, and as usual $x_{0}$ and $d$
are the mean position and width of the Gaussian respectively. The free
evolution of the Gaussian (\ref{gaussiennesslit}) is well-known to be given
(e.g., by applying the free propagator (\ref{amplitudelibre}) to Eq.
(\ref{gaussiennesslit})) by%
\begin{equation}%
\begin{split}
&  \psi_{{}}^{j}(x,t)=\bigg(\frac{2}{\pi}\bigg)^{\frac{1}{4}}\frac{e^{-i\pi
/4}}{\sqrt{\frac{\hbar t}{md}-2id}}\\
&  \exp\bigg\lbrace-\frac{md^{2}((\delta_{j}x_{0}-x)+p_{i}^{j}t/m)^{2}}%
{4d^{4}m^{2}+\hbar^{2}t^{2}}+i\frac{4d^{4}mp_{i}^{j}(2mx-p_{i}^{j}t)+\hbar
^{2}t\left(  m(x-x_{0})^{2}+2p_{i}^{j}x_{0}t\right)  }{8d^{4}hm^{2}+2\hbar
^{3}t^{2}}\bigg\rbrace.\label{evolcoherentslit1}%
\end{split}
\end{equation}
\newline The maximum of $\left\vert \psi_{{}}^{j}(x,t)\right\vert ^{2}$ thus
moves along the classical trajectory
\begin{equation}
x_{i}^{j}(t)=\delta_{j}x_{0}+\frac{p_{i}^{j}}{m}t. \label{class}%
\end{equation}
Note that if the Gaussian remains sufficiently narrow between the slits and
the screen (that is narrower than the spatial resolution of the probes or
screen detector), $\psi_{{}}^{j}(x,t)$ can be said to be carried effectively
by the single classical trajectory defined by Eq. (\ref{class}). In this case
the pre-selected state (\ref{initial}) involves two trajectories, with
$p_{i}^{1}$ and $p_{i}^{2}$ chosen so that these two trajectories meet at some
point, say $x_{1}$ on the detection screen (see Fig. \ref{idealpaths}).

\subsection{Post-selection}

For the post-selected wavefunction, the natural choice would be to select a
position $\left\vert x_{f}\right\rangle $ on the detection screen, or more
realistically a window function of the size of the spatial resolution. Rather
than dealing with unit-step functions, we will work for simplicity with a
Gaussian centered on $x_{f}$ and further assume we can control by collimation
the incoming mean momentum $p_{k}^{\prime}$ at the screen,%
\begin{equation}
\Xi_{x_{f},p_{k}^{\prime}}(x,t_{f})=\frac{1}{(2\pi\Delta^{2})^{\frac{1}{4}}%
}e^{i\frac{(x-x_{f})^{2}}{4\Delta^{2}}-i\frac{p_{k}^{\prime}(x-x_{f})}{h}}.
\label{gaussiennefin}%
\end{equation}
where the width $\Delta$ is of the order of the spatial resolution of the
detector. We will actually define for the controlled case studied here the
post-selected wave function to be given by a sum over such Gaussians all
centered on $x_{f}$
\begin{equation}
\chi_{x_{f}}(x,t_{f})=\sum_{k}c_{k}\Xi_{x_{f},p_{k}^{\prime}}(x,t_{f}),
\label{coherent2}%
\end{equation}
with weights $c_{k}$. The advantage of this choice is that the maximum of each
$\Xi_{x_{f},p_{k}^{\prime}}(x,t)$ at a backward evolved time $t$ can be seen
to move, similarly to Eq. (\ref{evolcoherentslit1}), along the classical trajectory%

\begin{equation}
x_{k}^{\prime}(t)=x_{f}+\frac{p_{j}^{\prime}}{m}(t_{f}-t). \label{backt}%
\end{equation}

\subsection{Weak trajectories}

\label{[sec-wt}

Assume a probe placed at $\mathbf{r}_{a}=(x_{a},z_{a})$ is weakly coupled to
the system with the interaction Hamiltonian given by Eq. (\ref{ham}).\ The
pre-selected state is given by Eq. (\ref{initial}) and let us take the
post-selected state to be given by Eq. (\ref{coherent2}) with a single term
($c_{1}=1,$ $c_{k\neq1}=0$). If the evolving pre-selected state remains
narrow, post-selection can only be successful if the post-selection position
and momentum collimation are set to%

\begin{equation}
x_{f}=\left\{
\begin{array}
[c]{c}%
x_{0}+p_{i}^{1}t_{f}/m\\
-x_{0}+p_{i}^{2}t_{f}/m
\end{array}
\right.  \qquad p_{k}^{\prime}=\left\{
\begin{array}
[c]{c}%
p_{i}^{1}\\
p_{i}^{2}%
\end{array}
\right.  . \label{condi}%
\end{equation}
From Eqs. (\ref{wv1})-(\ref{wv2}), we see that the weak value will be non-zero
only if the probe is positioned on the classical trajectory between the slit
and the post-selection point, since otherwise $\chi^{\ast}(x_{a})\psi
(x_{a})\approx0$. If we assume an array of probes arranged as in Fig.
\ref{idealpaths}, the same reasoning applies, now based on Eq. (\ref{idiprobe}%
): given a final point, say the point $x_{f}=x_{1}$ shown in Fig.
\ref{idealpaths}, each probe along the trajectory coming from slit 1 (blue
colored on Fig. \ref{idealpaths}) will be shifted if post-selection is chosen
with $p_{k}^{^{\prime}}=p_{i}^{1}.$ This set of shifted probes $\mathcal{M}%
_{\text{ab}}$ defines the weak trajectory given the pre and post-selected
states. Similarly, if the post-selected state is chosen so that $p_{k}%
^{^{\prime}}=p_{i}^{2}$, then only the probes along the red trajectory coming
from slit 2\ will be found to have shifted, defining the weak trajectory
obtained with this different post-selected state.

Finally, let us choose the post-selected state to be given by Eq.
(\ref{coherent2}) but now with $c_{1}=c_{2}=1/\sqrt{2},$ $c_{k\neq1,2}=0$ and
set $p_{1}^{^{\prime}}=p_{i}^{1},p_{2}^{^{\prime}}=p_{i}^{2}$. Now the set of
probes $\mathcal{M}_{\text{ab}}$ along the trajectories coming from both slits
(the blue and the red trajectories of Fig. \ref{idealpaths}) will be detected
to have shifted. With this choice of post-selection we have obtained a
superposition of weak trajectories, that could in principle be observed by
monitoring the probes $\mathcal{M}_{\text{ab}}$. This superposition of weak
trajectories can be understood as a sum over two effective Feynman paths.

\section{Weak trajectories in a double-slit setup\label{scheme}}

\subsection{Overlapping wavefronts}

A double slit setup typically displays wide interfering wavefronts expanding
from the slits to the screen instead of narrow wavepackets propagating along
one effective coarse-grained trajectory. Each wavefront is carried by several
paths emanating from the slits (recall here that due to the factorization
assumption of Sec. \ref{sec-2}, the wavefront in the far field moves at
constant speed $p_{z}/m$). Moreover several paths interfere at the location of
a probe. Identifying weak trajectories in this context from a grid of weakly
coupled probes is more involved than in the controlled case with narrow wavepackets.

In this situation the wavefunction in the Fraunhofer region or at the
detection screen is given by Eq. (\ref{kx2}), that can be shown to be
proportional to Eq. (\ref{evolcoherentslit1}) with the replacements
$t\rightarrow t-\tau$, $(\Delta x)^{2}\rightarrow2d^{2}+\hbar^{2}t^{2}%
/(2m^{2}d^{2})$ and $\tau\rightarrow m\delta_{j}x_{0}/p_{i}^{j}$. Hence we
will set the pre-selected wavefunction to be again given by Eq. (\ref{initial}%
) except that now neither the width nor the initial momentum are controlled,
but depend on the slit geometry and on the position of the source. We will
still assume however that we can control the post-selected state, by
collimating the mean momentum of the particle incoming on the screen at
$x_{f}$, and by choosing narrow window of detection (the parameter $\Delta$ in
Eq. (\ref{gaussiennefin})). Hence the post-selected wavefunction is identical
to Eq. (\ref{coherent2}).

\begin{figure}[h]
\centering \includegraphics[width=8.2cm]{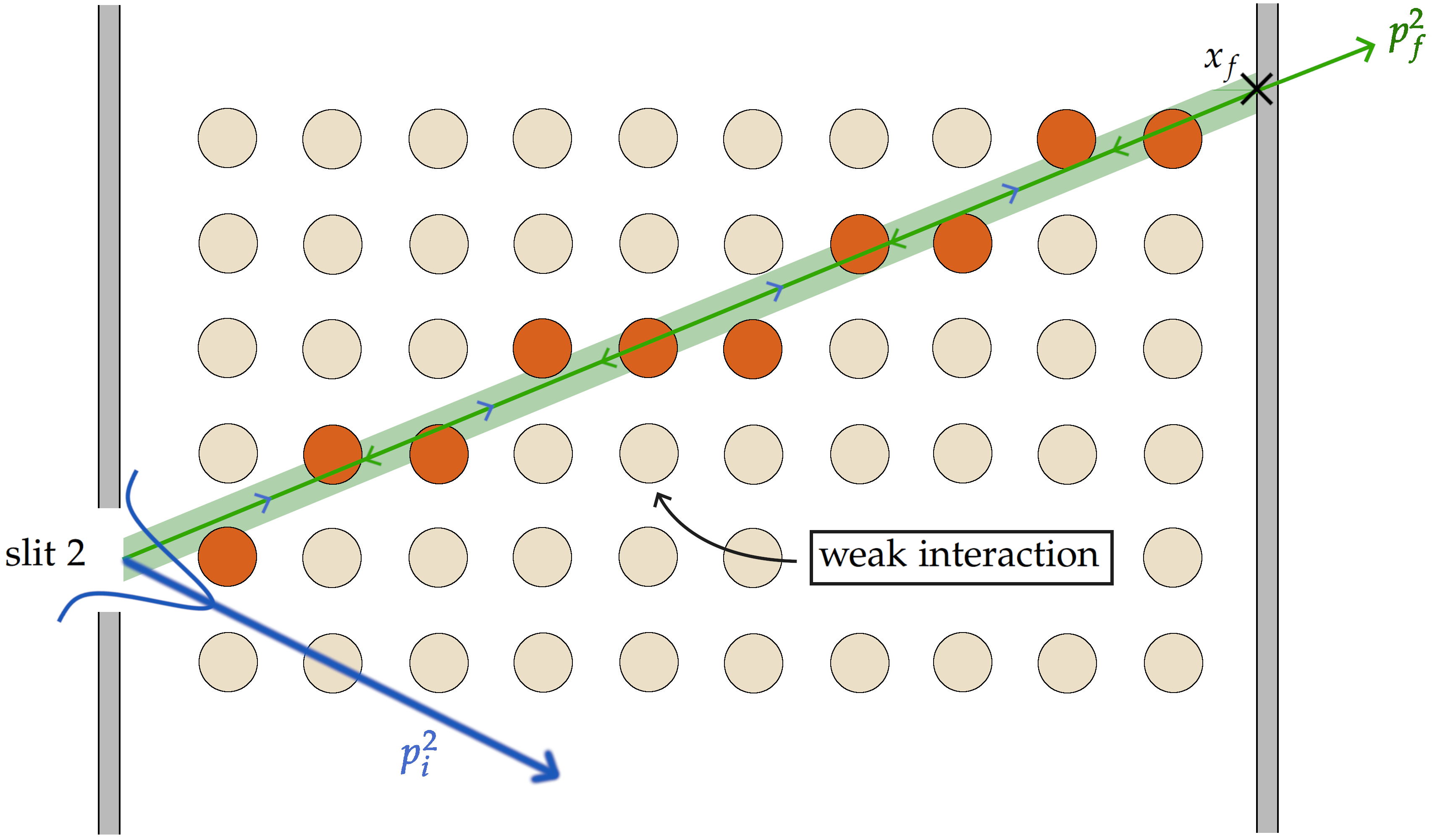}\caption{Schematic
representation (not to scale) of a grid of weakly coupled probes interacting
with the particle sent from a source (placed at $x=0$ as shown in Fig.
\ref{fentes}), hence without control of the transverse momentum. Here only
slit 2 is open and the post-selected state is collimated to filter along the
direction of the momentum $p_{2}^{f}$ indicated by the arrow. The probes
having shifted after post-selection are shown in darker tones. }%
\label{cas1}%
\end{figure}

\subsection{Weak measurement of a single classical trajectory}

\label{trajmoy} \noindent Assume only one slit is open, say slit 2. In this
case, the pre-selected is given by
\begin{equation}
\psi_{p_{i}}^{2}(x,t)=\frac{1}{(2\pi d^{2})^{\frac{1}{4}}}e^{ -i
\frac{(x+x_{0})^{2}}{4d^{2}}+ip_{i}(x+x_{0})/\hbar}%
\end{equation}
[cf. Eq. (\ref{gaussiennesslit})].\ Let us set the post-selected wavefunction
$\chi_{x_{f}}(x,t)=\Xi_{x_{f},p_{f}^{2}}(x,t)$ [cf. Eq. (\ref{coherent2})] to
be collimated along a single momentum $p_{f}^{2}$ chosen such that the
classical trajectory emerging from the center of slit 2 reaches $x_{f}$ at
time $t_{f}$, as shown on Fig. \ref{cas1}. Hence%
\begin{equation}
x_{f}^{1}(t_{f})=x_{0}+\frac{p_{f}^{1}}{m}t_{f}. \label{classical}%
\end{equation}
Let us place an array of probes $\mathcal{M}_{ab}$ as shown in Fig.
\ref{cas1}. The probes weakly measure the projectors $\hat{\Pi}_{a}$ at times
$t_{b}$. The weak values displayed by the weak detectors are given by
\begin{equation}
(\Pi_{a})_{w}(t_{b})=\frac{\chi_{x_{f}}^{\ast}(x_{a},t_{b})\psi_{p_{i}}%
^{2}(x_{a},t_{b})}{\int dx\chi_{x_{f}}^{\ast}(x,t_{b})\psi_{p_{i}}^{2}%
(x,t_{b})}, \label{wvj}%
\end{equation}
where $\chi_{x_{f}}^{\ast}(x,t)=\Xi_{x_{f},p_{f}^{2}}^{\ast}(x,t)$ is given by
Eq. (\ref{gaussiennefin}).

Sufficiently far from the slits (where the Fraunhofer approximation holds),
$\psi_{p_{i}}^{2}(x_{a},t_{b})$ will be non-zero, except accidentally if
$x_{a}$ happens to lie on a node of the wavefunction; we discard this
possibility since in practice the interaction does not take place at a single
point but over a region centered on $\mathbf{r}_{a}$ \footnote{In this case,
the numerator in Eq. (\ref{wvj}) should be replaced by $\int\chi_{x_{f}}%
^{\ast}(x^{\prime},t_{b})\psi_{p_{i}}^{2}(x^{\prime},t_{b})\Gamma
_{a}(x^{\prime})dx^{\prime}$ where $\Gamma_{a}(\mathbf{r})$ gives the spatial
profile of the interaction.\ For simplicity we have taken $\Gamma
_{a}(\mathbf{r})=\delta(\mathbf{r}-\mathbf{r}_{a})$ throughout the paper.}.
Therefore $(\Pi_{a})_{w}(t_{b})$ will be non-zero only if $\mathbf{r}_{a}$
lies near the trajectory defined by the post-selected momentum. If $p_{f}^{2}$
is chosen to be along the classical trajectory going from $-x_{0}$ (the center
of slit 2) to $x_{f}$ , then $(\Pi_{a})_{w}\neq0$ for all the probes
$\mathcal{M}_{ab}$ in the neighborhood of that trajectory. This case, pictured
in Fig. \ref{cas1} is similar to the cases discussed in Sec. \ref{secideal},
in which the weak trajectory links the slit to the post-selection point.

\begin{figure}[tb]
\centering \includegraphics[width=8.2cm]{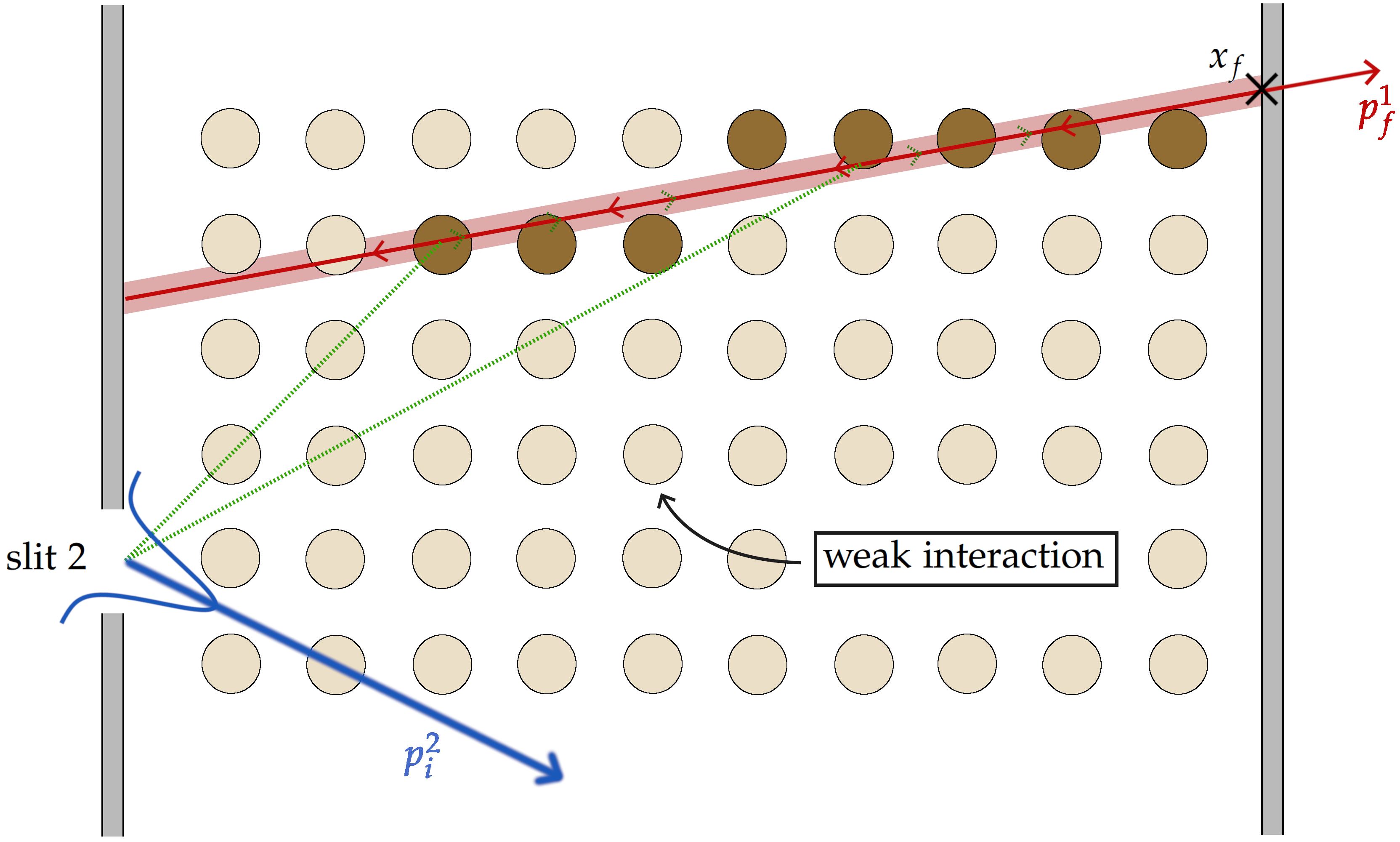}\caption{Same as Fig.
\ref{cas1} but with a different collimation direction. The resulting weak
trajectory only appears in the Fraunhofer region (see text for details). }%
\label{cas2}%
\end{figure}

However, we may pick $p_{f}^{2}$ along a different direction, as portrayed in
Fig. \ref{cas2}. In this case, we know from the propagator [Eq.
(\ref{amplitudelibre})] that $\chi_{x_{f}}^{\ast}(x_{a},t_{b})$ will be
non-zero essentially along the backward evolved trajectory given by Eq.
(\ref{backt}). The probes $\mathcal{M}_{ab}$ for values of $t_{b}$
sufficiently large (that is far from the slits, where the wavefront is
approximately uniform along $z$) will therefore shift, but not those closer to
the slits since for small values of $t_{b}$ we have $\psi_{p_{i}}^{2}%
(x_{a},t_{b})\approx0$ if $x_{a}$ is appreciably different from $-x_{0}$. Such
a situation gives rise to the ``incomplete'' weak trajectory shown in Fig.
\ref{cas2}.

\begin{figure}[tb]
\centering \includegraphics[width=8.2cm]{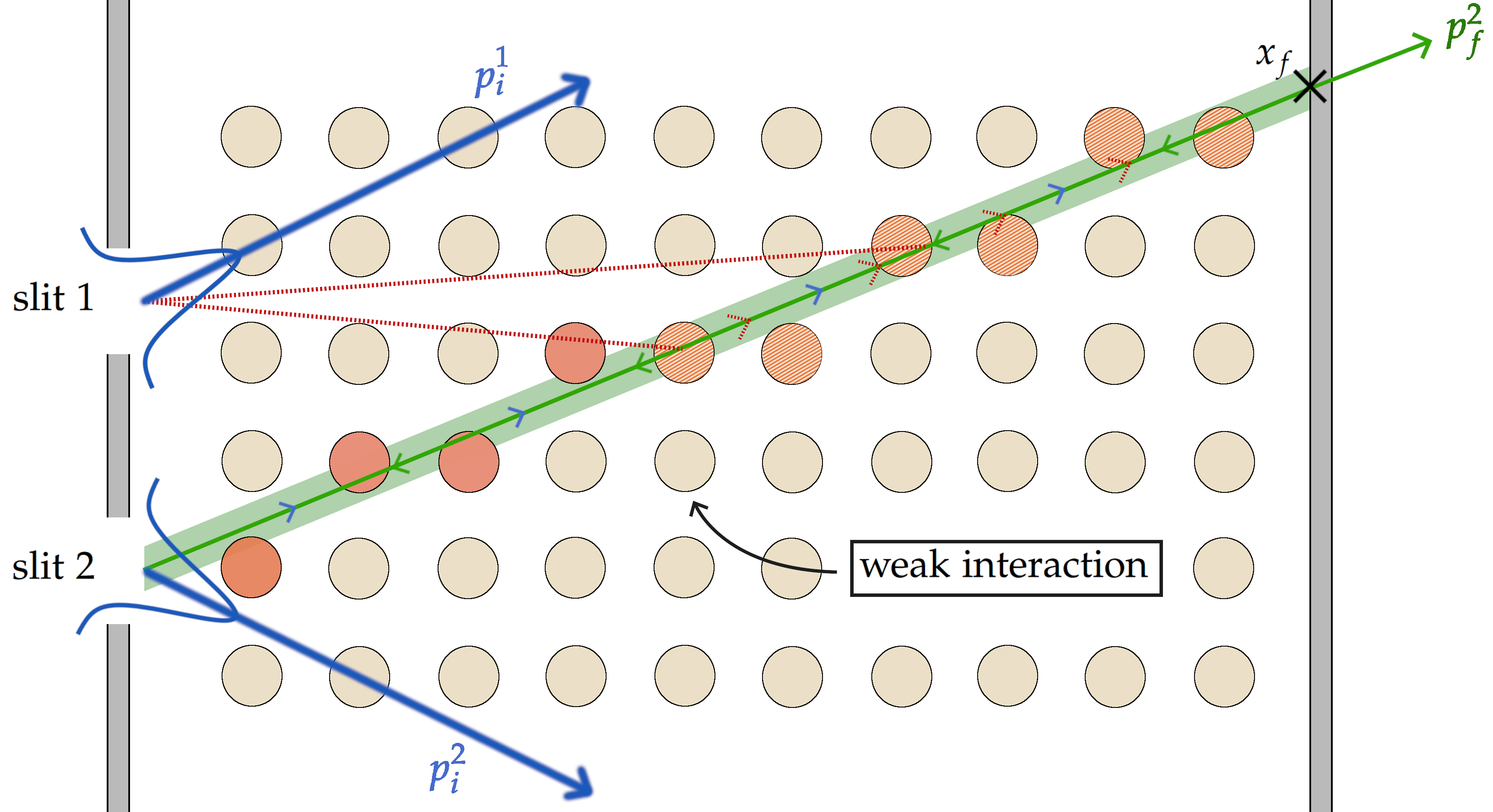}\caption{Same as Fig.
\ref{cas1} but with both slits open. The shifts of some probes in the
Fraunhofer region change due to the paths coming from slit 1 (they are
pictured with a lighter tone), though the weak trajectory is the same as in
Fig. \ref{cas1}. }%
\label{cas3}%
\end{figure}

Finally, let us consider the 2-slit situation with the same post-selected
state. Recalling that the initial wavefunction is given by Eq. (\ref{initial}%
), the weak value (\ref{wvj}) resulting from the coupling of probe
$\mathcal{M}_{a,b}$ now becomes
\begin{equation}
(\Pi_{a})^{w}(t_{b})=\frac{\chi_{x_{f}}^{\ast}(x_{a},t_{b})\left(  \psi
_{p_{i}^{1}}^{1}(x_{a},t_{b})+\psi_{p_{i}^{2}}^{2}(x_{a},t_{b})\right)  }{\int
dx\chi_{x_{f}}^{\ast}(x,t_{b})\left(  \psi_{p_{i}^{1}}^{1}(x,t_{b}%
)+\psi_{p_{i}^{2}}^{2}(x,t_{b})\right)  }. \label{wvj2}%
\end{equation}

\begin{figure}[tb]
\centering \includegraphics[width=8.2cm]{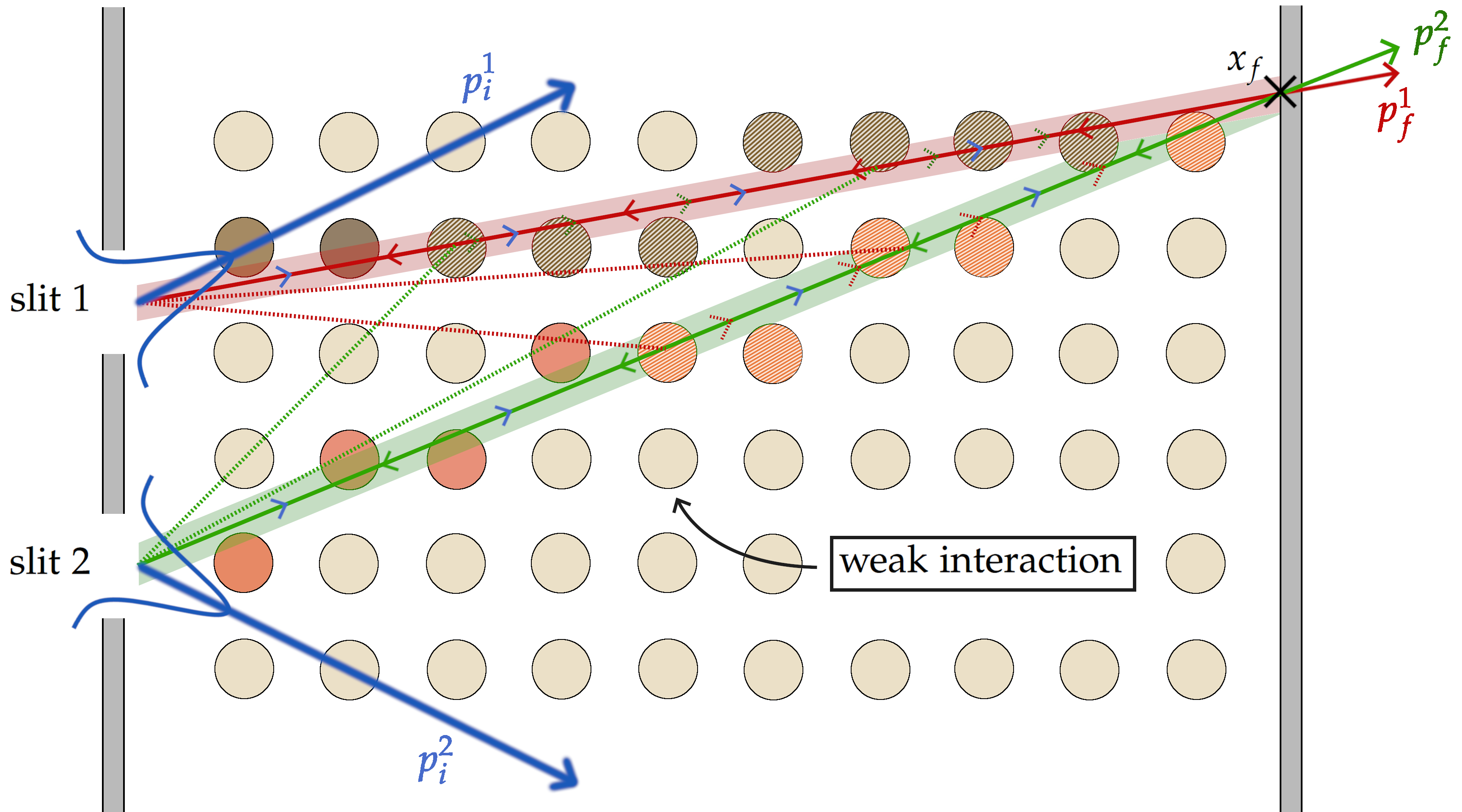}\caption{Same as Fig.
\ref{cas3} but now the post-selection filters along the directions $p_{f}^{1}$
and $p_{f}^{2}$ as indicated by the arrows pictured at $x_f$. The paths
superposition is identical to the ideal case shown in Fig. \ref{idealpaths}
but here paths from both slits contribute to the weak value of most probes. }%
\label{cas4}%
\end{figure}
Hence opening the second slit does not affect dramatically the
weak trajectories -- for instance for the choice of $p_{f}^{2}$ corresponding
to Fig. \ref{cas1}, the weak trajectory does not change at all.\ However the
weak values do change, due to the contribution of the Feynman paths coming
from the other slit (see Fig. \ref{cas3}). The difference between the weak
values (\ref{wvj}) and (\ref{wvj2}) can be understood as a signature of the
interference (carried by additional paths) between the waves coming from each slit.

\subsection{Weak measurement of several classical trajectories\label{several}}

One can measure several weak trajectories with a single open slit by employing
a post-selected state of the form (\ref{coherent2}), similarly to the case
treated in Sec. \ref{secideal}. The difference here is that most weak probes
will receive contributions from paths coming from each slit, as in the case
described by Eq. (\ref{wvj2}). Let us take again the post-selected state to be
of the form given by Eq. (\ref{coherent2}),%
\begin{equation}
\chi_{x_{f}}(x,t)=\frac{1}{\sqrt{2}}\left(  \Xi_{x_{f},p_{f}^{1}}%
(x,t)+\Xi_{x_{f},p_{f}^{2}}(x,t)\right)  \label{defi99}%
\end{equation}
where $\Xi_{x_{f},p_{f}^{1}}$ is defined by Eq. (\ref{gaussiennefin}). The
resulting weak values are given by replacing $\chi_{x_{f}}(x_{a},t_{b})$ with
Eq. (\ref{defi99}) in Eq. (\ref{wvj2}). This situation is portrayed in Fig.
\ref{cas4}, in which $p_{f}^{1}$ and $p_{f}^{2}$ are chosen along the
trajectories linking the slits 1 and 2 to $x_{f}$. The weak probes that will
shift, defining the weak trajectories, are those between the slits and the
detector at $\mathbf{r}_{f}.$

The difference with the ideal case discussed in Sec.\ \ref{secideal} is that
now trajectories from both slits reach each probe. This gives yet another
signature, in terms of the weak values of the interference between the waves
coming from each slit. These waves are carried by Feynman paths propagating
along classical trajectories.\ For each weakly coupled probe, the weak value
depends on 3 trajectories (or 3 families of trajectories if one takes into
account the finite width of the weak probes and the screen detector): the
trajectories from each slit to the probe and the trajectory from the probe to
the post-selected detector on the screen. If one slit is closed, $\psi^{1}$ or
$\psi^{2}$ in Eq. (\ref{wvj2}) vanishes. This changes the weak values of the
weak probes in the Fraunhofer region.\ Note that the weak probes closer to the
shut slit will have null weak values, since both $\psi^{1}$ and $\psi^{2}$
will vanish at the location of the weak probes.

\begin{figure}[tb]

	\centering \includegraphics[width=8.2cm]{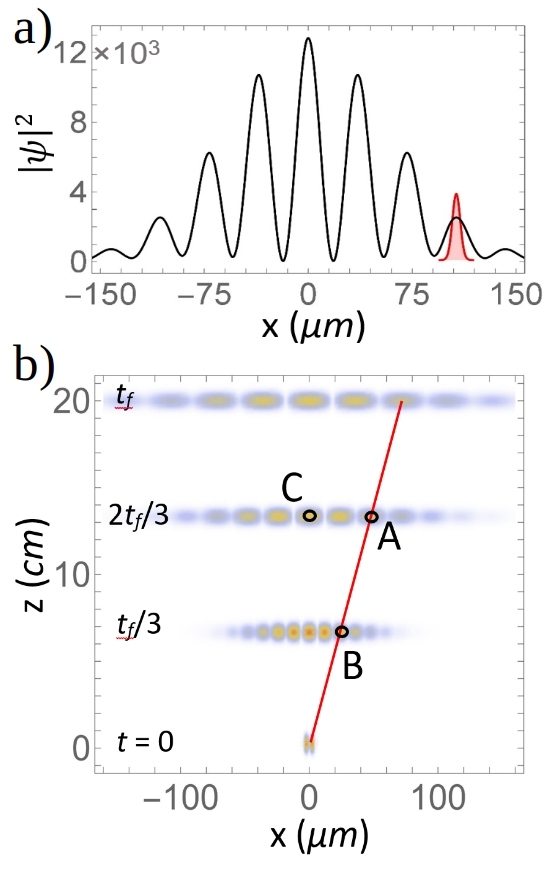}
	\caption{	 (a): interference pattern due to
		electron diffraction
		on a detection plane placed 20 cm from the slits (see text for the parameter values). The
		post-selected state is represented as a filled peak (color online: in red). (b) Snapshots of the electron density
		as the electron evolves from the slits (at $t=0$) to the detector (at $t=t_f$).
		The chosen post-selected state filters the wavefunction traveling along the red trajectory 
		going through the points A and B. The weak values at the positions A, B and C are given in Table \ref{tabwv}.\label{fignums} }%
\end{figure}

\subsection{Numerical illustration}

We present here an illustration with numbers corresponding to current typical
single electron diffraction experiments, see e.g. Ref. \cite{batelaan}. In
such experiments the slit plane dimensions is in the mesoscopic range, while
the interference figure is of macroscopic size.\ This leaves ample room to
insert pointers between the plane and detector planes, though discriminating
the post-selection between $p_{f}^{1}$ and $p_{f}^{2}$ as in Eq.
(\ref{defi99}) becomes unpractical, since $p_{f}^{1}\approx p_{f}^{2}$.

Fig. \ref{fignums}(a) shows the interference pattern on a detection screen
placed 20 cm from the slit plane (we thus have [refer to Fig. \ref{fentes} and
Eqs. (\ref{gaussiennesslit})-(\ref{class})] $m=m_{e}\approx9.1\times10^{-31}$
kg, $D=0.2$ m, and the other parameter values are $x_{0}=2$ $\mu$m, $d=0.2$
$\mu$m, $p_{i}^{1}/m=-p_{i}^{2}/m=100$ m/s, $p_{z}/m=10^{6}$ m/s), obtained by
evolving Eq. (\ref{initial}) up tp $t_{f}=mD/p_{z}=0.2$ $\mu$s. We also show
in Fig. \ref{fignums}(a) the post-selected state (\ref{gaussiennefin}) at
$t=t_{f}$, choosing $x_{f}=109.9$ $\mu$m at the maximum of the third peak to
the right of the central peak of the diffraction figure. The post-selected
momentum is chosen along the direction of slit 1, $p_{f}^{1}=m(x_{f}%
-x_{0})/t_{f}$. We set $\Delta=2$ $\mu$m.

Fig. \ref{fignums}(b) shows a density plot of the electron wavefunction as the
electron evolves from the slits to the detection screen. We also indicate
three positions A,B and C for which we compute weak values given in Table
\ref{tabwv}. The weak values are computed using an expression of the type
given by Eq. (\ref{wvj}) when only one slit is open, or Eq. (\ref{wvj2}) when
both slits are open. We integrate the numerator in Eqs. (\ref{wvj}%
)-(\ref{wvj2}) with the same smooth Gaussian [31] to take into account the
finite width of the coupling. The results for the real part of the weak values
are shown in Table \ref{tabwv}.\ It can be seen that the weak values are
markedly different depending on which slit is open.\ The last column (for a
weak pointer C) is meant to illustrate that any weak value will vanish if it
does not lie along the trajectory defined by the post-selected state (the red
line in Fig. \ref{fignums}(b)).
\begin{table}
\begin{tabular}
[c]{|c|c|c|c|}\hline
& A ($t=2t_{f}/3)$ & B ($t=t_{f}/3)$ & C (any $t$)\\\hline
Both slits open & 0.38 & 4.26 & 0\\\hline
Slit 1 ($x_{0}$) open only & -1.79 & 8.18 & 0\\\hline
Slit 2 (-$x_{0}$) open only & 0.75 & 1.51 & 0\\\hline
\end{tabular}

\caption{Real part of the spatial projector weak value at points A, B and C shown in Fig. \ref{fignums}(b). Different cases with slits open or closed are considered, giving rise to different weak values.}
\label{tabwv}
\end{table}

\section{Simplified protocol for a photonic implementation\label{photonic}}

\subsection{Overview}

In this Section, we detail a concrete protocol to measure weak trajectories
having in mind a photonic setup, in the line of recent experiments dealing
with single photons going through slits \cite{urbasi,kocsis}. The protocol is
a simplified version of the double-slit setup case seen in the previous
section (Sec. \ref{scheme}).

The main difference with the previous setups is that the weak probes are not
independent systems that couple to the photon, but optical elements that
create a coupling between the spatial degree of freedom of the photon and its
polarization. The result of the coupling is encoded in the photon itself.
Hence it is not possible to imagine a dense array of probes whose state would
be read out after post-selection. Here in order to extract the rotations due
to the weak coupling with the optical elements, we need to implement a
protocol with a small number of glass plates or birefringent crystals placed
between the slits and the detector screen. In the present protocol, we will
work with birefringent crystals coupling at a given position the transverse
momentum along $x$, $\hat{k}$ to the photon polarization. We will therefore
deal with the corresponding weak values $\left(  k\right)  ^{w}$ of the
transverse momentum, instead of the weak values of the position $\hat{\Pi}%
_{a}$.

Nevertheless, we will see that the weak measurement of $\hat{k}$ allows us to
observe the superposition of trajectories. The rationale is that a non-null
weak value implies that the photonic wave has interacted with the birefringent
crystal.\ The weak probe is now the photon polarization, which is rotated
(rather than shifted) by a quantity proportional to $\left(  k\right)  ^{w}%
.$\ Reading out the polarization at post-selection allows us to infer the past
interactions of the photon with the birefringent crystals, and from there to
reconstruct the weak trajectories.

\subsection{Interactions and weak values}

\begin{figure}[h]
\centering \includegraphics[width=8.4cm]{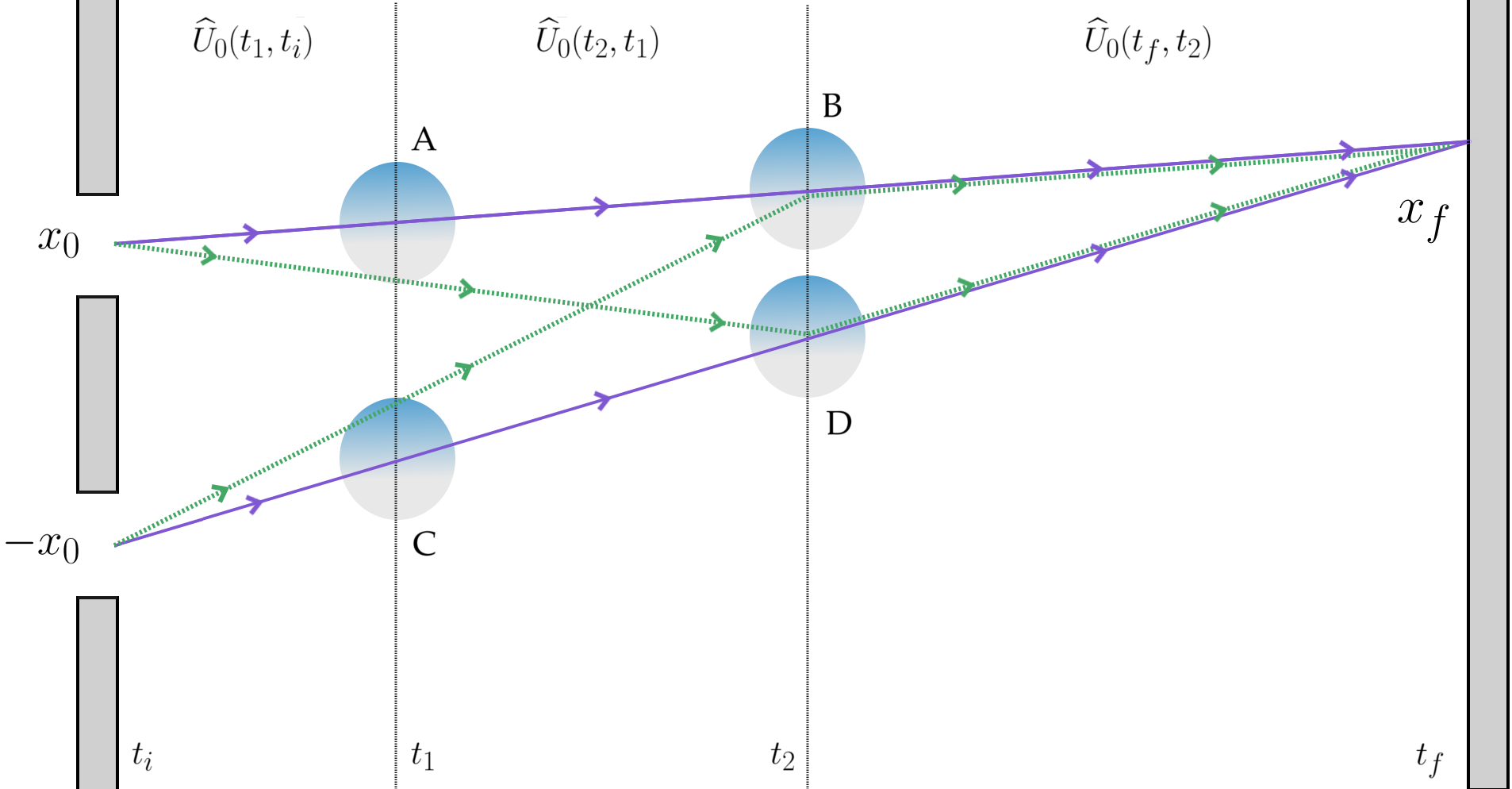}\caption{Schematic diagram
(not to scale) showing 4 birefringent crystals creating a weak coupling in
order to measure the paths superposition in a photonic setup. A and B are
disposed near the slits while B and D are located in the Fraunhofer region.
The paths joining the slits to the crystals and to $x_{f}$ (the post-selected
position on the detection screen) are the weak trajectories inferred from the
measurement of the polarization intensities.}%
\label{fentessetup}%
\end{figure}

Consider the setup represented in Fig. \ref{fentessetup}. Two of the
birefringent crystals (A and C) are near the slits (in the region with
spherical wavefronts) while B and D are in the Fraunhofer region. The
pre-selected state is again of the form given by Eqs. (\ref{initial}%
)-(\ref{gaussiennesslit}), where the mean momentum of each Gaussian is not
controlled but depends on the relative positions of the source and the slits
plane, as in Sec. \ref{scheme}. The initial polarization is chosen to be
diagonal, $\left\vert \nearrow\right\rangle =\left(  \left\vert H\right\rangle
+\left\vert V\right\rangle \right)  /\sqrt{2},$ so the photon state at the
slits is (setting $t=0$ when the photon wavefunction crosses the slits)%
\begin{equation}
\frac{1}{\sqrt{2}}\left(  \psi_{p_{i}^{1}}^{1}(x,0)+\psi_{p_{i}^{2}}%
^{2}(x,0)\right)  \left\vert \nearrow\right\rangle . \label{inistate}%
\end{equation}
The interaction Hamiltonian between the photon wavefunction and a birefringent
crystal is given by
\begin{equation}
\hat{H}_{a}=g_{a}\hat{k}\hat{\sigma}_{\alpha} \label{hama}%
\end{equation}
where $g_{a}$ is the coupling constant and $\hat{\sigma}_{\alpha}=\frac{\hbar
}{2}\hat{\Pi}_{a}\sigma_{3}$ is the linear polarization observable at the
position $\left(  x_{a},z_{a}\right)  $ of the crystal ($\sigma_{3}=\left\vert
H\right\rangle \left\langle H\right\vert -\left\vert V\right\rangle
\left\langle V\right\vert $ is the third Pauli matrix). Finally we will use
the post-selected state $\left\vert \chi_{x_{f}}\right\rangle $ employed
above, defined by Eq. (\ref{defi99}), that is a superposition of 2 narrow
Gaussians centered on $x_{f}$ and with the momentum collimated along the
directions joining $\mathbf{r}_{f}$ to the center of the slits.

In the weak coupling limit, each interaction (\ref{hama}) gives rise to the
weak value
\begin{equation}
(k_{a})^{w}(t_{b})=\frac{\chi_{x_{f}}^{\ast}(x_{a},t_{b})(-i\hbar
\partial_{x_{a}})\left(  \psi_{p_{i}^{1}}^{1}(x_{a},t_{b})+\psi_{p_{i}^{2}%
}^{2}(x_{a},t_{b})\right)  }{\int dx\chi_{x_{f}}^{\ast}(x,t_{b})\left(
\psi_{p_{i}^{1}}^{1}(x,t_{b})+\psi_{p_{i}^{2}}^{2}(x,t_{b})\right)  },
\end{equation}
similarly to Eq. (\ref{wvj2}). As we remarked above, the numerator of this
expression should more realistically be integrated over the width $\Gamma(x)$
of the crystal. The important point here is that the polarisation is rotated
as a result of this interaction.\ For instance the interaction of a crystal at
$\mathbf{r}_{a}$ with the photon in state $\left\vert \psi(t)\right\rangle
\left\vert \nearrow\right\rangle $ for $t<t_{b}$ results after post-selection
in changing the polarisation to
\begin{equation}
\left\vert P\right\rangle =e^{-i\gamma_{a}(k_{a})^{w}}\left\vert
H\right\rangle +e^{i\gamma_{a}(k_{a})^{w}}\left\vert V\right\rangle \text{,}%
\end{equation}
where $\left\vert P\right\rangle $ stands here for the polarisation state and
where $\gamma_{a}=\int dtg_{a}/\hbar$ . At post-selection, this polarisation
state is read-out by projecting $\left\vert P\right\rangle $ in a chosen
basis; here we will employ the diagonal basis $\{|\nearrow\rangle
,|\swarrow\rangle\}$. The polarisation state thus encodes, similarly to the
weak probe shifts in the ideal cases mentioned earlier, the past interactions
with specific crystals. However, the difference is that in the ideal case with
independent weak probes disposed on a grid, one could in principle read the
shift indicated by each weak probe. Here, the rotation due to the interaction
with each crystal cannot be read-out since there are several crystals, each
inducing a different rotation. The protocol we now detail circumvents this
problem by introducing several steps in order to extract the weak rotations
from the observed polarisation intensities.

\subsection{Protocol\label{protox}}

In order to extract the weak values from the polarisation intensities detected
at post-selection, we will introduce our protocol based on the setup shown in
Fig. \ref{fentessetup} involving 4 birefringent crystals. For definiteness we
assume the initial and post-selected states to be given by Eqs.
(\ref{inistate}) and (\ref{defi99}) respectively. The protocol is not unique
and its main idea can easily be adapted to other configurations.

Each of the 4 crystals interacts with the photon through a Hamiltonian given
by\ Eq. (\ref{hama}), that we will write with an obvious notation $\hat{H}%
_{A},\hat{H}_{B}$, etc... A (resp. C) is disposed close enough to slit 1
(resp.\ slit 2) so that the wave coming from the other slit has negligible
amplitude. B\ and C\ are reached by waves coming from both slits. The idea
underlying the protocol is to infer the rotation due to the interaction with
each optical element. To do so, in a first step we place the crystal A only,
so that $\hat{H}_{A}$ only comes into play. In a second step, we also place
the crystal C and the resulting interaction Hamiltonian is $\hat{H}_{A}%
+\hat{H}_{C}$. In a third step, we insert in addition the crystals B and D.
Finally, in a fourth step we place a phase-shifter on crystal B, in order to
create a $\pi$ shift.\ 

Let us denote by $|P_{n}\rangle$ the final polarization state after the
realization of the $n$th step. We have
\begin{align}
|P_{1}\rangle &  =\zeta\big(e^{-i\gamma_{A}(k_{A})^{w}}|H\rangle
+e^{i\gamma_{A}(k_{A})^{w}}|V\rangle\big)\\
|P_{2}\rangle &  =\zeta\big(e^{-i\gamma_{A}(k_{A})^{w}-i\gamma_{C}(k_{C})^{w}%
}|H\rangle+e^{i\gamma_{A}(k_{A})^{w}+i\gamma_{C}(k_{C})^{w}}|V\rangle\big)\\
|P_{3}\rangle &  =\zeta\big(e^{-i\gamma_{A}(k_{A})^{w}-i\gamma_{C}(k_{C}%
)^{w}-i\gamma_{B}(k_{B})^{w}-i\gamma_{D}(k_{D})^{w}}|H\rangle+e^{i\gamma
_{A}(k_{A})^{w}+i\gamma_{C}(k_{C})^{w}+i\gamma_{B}(k_{B})^{w}+i\gamma
_{D}(k_{D})^{w}}|V\rangle\big)\\
|P_{4}\rangle &  =\zeta\big(e^{-i\gamma_{A}(k_{A})^{w}-i\gamma_{C}(k_{C}%
)^{w}+i\gamma_{B}(k_{B})^{w}-i\gamma_{D}(k_{D})^{w}}|H\rangle+e^{i\gamma
_{A}(k_{A})^{w}+i\gamma_{C}(k_{C})^{w}-i\gamma_{B}(k_{B})^{w}+i\gamma
_{D}(k_{B})^{w}}|V\rangle\big),
\end{align}
where $\zeta=\langle\chi_{x_{f}}(t_{f})|\psi(t_{f})\rangle/\sqrt{2}$. For each
step the intensities in the diagonal basis, $I_{n}^{\nearrow}=|\langle
\nearrow|P_{n}\rangle|^{2}$ and $I_{n}^{\swarrow}=|\langle\swarrow
|P_{n}\rangle|^{2}$ are measured. From these intensities we can recover the
rotations (or the weak values if the effective couplings $\gamma_{a}$ are
known, e.g. from calibration). Indeed, introducing the contrast $C_{n}$ as
\begin{equation}
C_{n}=\frac{I_{n}^{\nearrow}-I_{n}^{\swarrow}}{I_{n}^{\nearrow}+I_{n}%
^{\swarrow}}, \label{cont}%
\end{equation}
the weak values $(k_{A})^{w}$ and $(k_{C})^{w}$ can be inferred from
\begin{align}
\gamma_{A}(k_{A})^{w}  &  =\frac{1}{2}\arccos(C_{1})\qquad\label{kaw}\\
\gamma_{C}(k_{C})^{w}  &  =\frac{1}{2}\arccos(C_{2})-\gamma_{A}(k_{A})^{w},
\end{align}
whereas the weak values $(k_{B})^{w}$ and $(k_{D})^{w}$ are deduced from
\begin{align}
\gamma_{B}(k_{B})^{w}  &  =\frac{1}{2}(\arccos(C_{3})-\arccos(C_{4}))\\
\gamma_{D}(k_{D})^{w}  &  =\frac{1}{2}(\arccos(C_{3})+\arccos(C_{4}%
))-2\gamma_{A}(k_{A})^{w}-2\gamma_{C}(k_{C})^{w}. \label{kdw}%
\end{align}

The knowledge of the weak values is not sufficient to identify the weak
trajectories, because several trajectories (or families of trajectories) may
contribute to a given weak value. In the present setup, the crystals A and C
receive a wave from a single slit, essentially emanating from the center of
slits 1\ and 2 respectively. So observing non-vanishing weak values
$(k_{A})^{w}$ and $(k_{C})^{w}$ is a signature of the paths linking each slit
to $\mathbf{r}_{f}$ (see the purple paths in Fig. \ref{fentessetup}) . For the
crystals B and D however we are in the situation described in Sec.
\ref{several}: as pictured in Fig. \ref{fentessetup}, each of these crystals
receives a wave from each slit, a wave carried by the path from the center of
the slit to the crystal (the neigboring paths also contribute, but the path at
the slit center has the highest amplitude). In order to parse the contribution
coming from each slit, we write%
\begin{align}
(k_{B})^{w}  &  =(k_{B}^{11})^{w}+(k_{B}^{12})^{w}\label{b}\\
(k_{D})_{w}  &  =(k_{D}^{22})^{w}+(k_{D}^{21})^{w} \label{d}%
\end{align}
where%
\begin{equation}
(k_{a}^{jl})^{w}=\frac{\Xi_{x_{f},p_{j}^{}}^{\ast}(x_{a},t_{b})(-i\hbar
\partial_{x_{a}})\psi_{p_{i}^{l}}^{l}(x_{a},t_{b})}{\sqrt{2}\int dx\chi
_{x_{f}}^{\ast}(x,t_{b})\left(  \psi_{p_{i}^{1}}^{1}(x,t_{b})+\psi_{p_{i}^{2}%
}^{2}(x,t_{b})\right)  }.
\end{equation}
We have assumed in writing Eqs. (\ref{b}) and (\ref{d}) that each Gaussian
$\Xi_{x_{f},p_{j}^{}}(x_{f},t_{f})$ of the post-selected state $\chi_{x_{f}}$
remains narrow so that $\Xi_{x_{f},p_{1}^{}}$ only filters the waves coming
from B and $\Xi_{x_{f},p_{2}^{}}$ only filters the waves coming from D. Under
these conditions, we see that the rotation induced by $(k_{B}^{11})^{w}$ is a
signature of the \textquotedblleft direct\textquotedblright\ path going from
slit 1 to $\mathbf{r}_{f}$ (represented in purple in Fig. \ref{fentessetup})
whereas $(k_{B}^{12})^{w}$ is due to the path going from slit 2 to B and from
B to $\mathbf{r}_{f}$ (shown in green in Fig. \ref{fentessetup}). Similarly
$(k_{D}^{22})^{w}$ is a signature of the path linking slit 2 to $\mathbf{r}%
_{f}$ while $(k_{D}^{21})^{w}$ singles out the path slit 1 $\rightarrow$ D
$\rightarrow\mathbf{r}_{f}$.

In order to extract the quantities defined in Eqs. (\ref{b})-(\ref{d}) from
experimental observations, we need the intensities with one or the other slit
closed. Doing so changes the pre-selected state to either $\psi_{p_{i}^{1}%
}^{1}(x,t=0)$ or $\psi_{p_{i}^{2}}^{2}(x,t=0)$. Assume only slit $l=1,2$ is
open. The weak values due to the weak couplings at B and D, denoted
$\kappa_{B}^{l}$ and $\kappa_{D}^{l}$ are given by%

\begin{align}
\kappa_{B}^{l}  &  =\frac{\Xi_{x_{f},p_{1}^{}}^{\ast}(x_{B},t_{B}%
)(-i\hbar\partial_{x_{B}})\psi_{p_{i}^{l}}^{l}(x_{B},t_{B})}{\int
dx\chi_{x_{f}}^{\ast}(x,t_{B})\psi_{p_{i}^{l}}^{l}(x,t_{B})}\label{kappab}\\
\kappa_{D}^{l}  &  =\frac{\Xi_{x_{f},p_{2}^{}}^{\ast}(x_{D},t_{D}%
)(-i\hbar\partial_{x_{D}})\psi_{p_{i}^{l}}^{l}(x_{D},t_{B})}{\int
dx\chi_{x_{f}}^{\ast}(x,t_{B})\psi_{p_{i}^{l}}^{l}(x,t_{B})}. \label{kappad}%
\end{align}
These single slit weak values $\kappa_{a}^{l}$ are obtained from intensity
measurements similarly to the method employed in order to observe the weak
values $(k_{a})^{w}$ (see details in the Appendix).

We now note that the weak values $\kappa_{a}^{l}$ can be written in terms of
the quantities defined in the right handside of Eqs. (\ref{b}) and
(\ref{d}).\ For example%
\begin{equation}
(k_{B})^{w}=\kappa_{B}^{1}\frac{\langle\chi(t_{f})|\psi_{p_{i}^{1}}^{1}%
(t_{f})\rangle}{\langle\chi(t_{f})|\psi(t_{f})\rangle}+\kappa_{B}^{2}%
\frac{\langle\chi(t_{f})|\psi_{p_{i}^{2}}^{2}(t_{f})\rangle}{\langle\chi
(t_{f})|\psi(t_{f})\rangle}; \label{uk}%
\end{equation}
the same expression holds for $(k_{D})^{w}$, so that the amplitude ratios in
Eq. (\ref{uk}) can be obtained from the independent measurements of
$(k_{B})^{w}$, $(k_{D})^{w},$ $\kappa_{B}^{1},$ $\kappa_{B}^{2}$, $\kappa
_{D}^{1}$ and $\kappa_{D}^{2}$. From there the terms in Eqs. (\ref{b}) and
(\ref{d}) are recovered, e.g. $(k_{B}^{11})^{w}=\kappa_{B}^{1}\left(
\langle\chi(t_{f})|\psi_{p_{i}}^{1}(t_{f})\rangle/\langle\chi(t_{f}%
)|\psi(t_{f})\rangle\right)  .$

\subsection{Weak trajectories from the observed weak values}

The protocol combines weak values extracted from different setups, aimed at
compensating the fact that rather than dealing with an array of weakly coupled
probes, a realistic photonic experiment would rely on coupling the transverse
momentum of the photon to its polarization when the photon interacts with a
birefringent crystal.

When only one slit is open, say slit 1, the classical path linking the slit to
the detection point $\mathbf{r}_{f}$ goes though A and B.\ This is captured by
the interaction of the photon with the crystals placed at A and B through the
rotation of the photon polarization encapsulated in the weak values $\left(
k_{A}\right)  ^{w}$ and $\kappa_{B}^{1}$. When slit 2 is open, $\left(
k_{A}\right)  ^{w}$ does not change, but $\kappa_{B}^{1}$ is modified to
$\left(  k_{B}\right)  ^{w}$: there is indeed an additional path taken by the
photon reaching B from the source through slit 2. This \textquotedblleft sum
over paths\textquotedblright\ appears here by parsing $\left(  k_{B}\right)
^{w}$ as the coherent superposition given by Eq. (\ref{b}). And of course,
there is also an additional path reaching the screen at $\mathbf{r}_{f}$
through slit 2, points C and D.\ This weak trajectory is evidenced by $\left(
k_{C}\right)  ^{w}$ (that becomes non-zero as slit 2\ is opened) and by the
change of the polarization due to the interaction of the photon with the
crystal placed at D (as a result the weak value changes from $\kappa_{D}^{1}$
to $\left(  k_{D}\right)  ^{w}$). Finally, if we close slit 1, the
corresponding weak trajectories disappear, and this is evidenced by the the
change in the polarization detected at post-selection ($\left(  k_{A}\right)
^{w}$ becomes 0, $\left(  k_{B}\right)  ^{w}$ and $\left(  k_{D}\right)  ^{w}$
become $\kappa_{B}^{2}$ and $\kappa_{D}^{2}$, while $\left(  k_{C}\right)
^{w}$ remains the same).

\section{Discussion and conclusion \label{sec-conc}}

The quantum formalism explains interferences in Young's double-slit setup by
asserting the particle's wavefunction goes through both slits. In Feynman's
path integral formulation, the waves that interfere are ``carried'' by
classical trajectories, so that the interference figure is seen as due to a
sum over paths. We have discussed how this path superposition could in
principle be observed by implementing weak measurements between the particle
and an array of weakly coupled probes, provided the particle can be
appropriately post-selected. The shifted probes reveal a coarse-grained form
of the sum over paths in terms of ``weak trajectories''. We further proposed a
concrete protocol in order to observe weak trajectories in a simple photonic
double-slit setup containing a small number of crystals placed between the
slits and the detector plane.

It should be emphasized that the specific form chosen for the weak measurement
process is instrumental in sampling different aspects of the dynamics. A weak
measurement of the momentum followed immediately by a projective position
measurement leads to the velocity field associated with the Schr\"{o}dinger
current density. From there it is possible to reconstruct Bohmian trajectories
\cite{leavens}, though as discussed elsewhere \cite{JPA,hiley} Bohmian
trajectories are not the observed quantities -- contrary to the paths measured
in this work that can be measured by following the wavefunction in space
(through the position $\mathbf{r}_{a}$ of the probes) and time (by turning on
and off the interaction at the desired time $t_{b}$). Indeed, in the case
examined here, a weak interaction is not followed immediately by
post-selection (that would terminate the system's evolution), but it is
followed by successive weak interactions, in order to sample the intermediate
dynamics before detecting the particle on the screen. Note that strictly
speaking photons do not admit Bohmian trajectories, so that single photon
trajectories reconstructed from experimentally obtained \cite{kocsis} weak values
 should be more appropriately related to the Poynting energy flow
\cite{addi}.

Another crucial aspect is the choice of post-selection: the post-selected
wavefunction filters the dynamical aspects that can be accessed at an
intermediate time.\ This aspect appeared in our scheme through collimation
along one or two directions upon post-selection (in order to filter the waves
coming from a specific direction).\ It is also possible to consider
post-selecting coherently and simultaneously in distinct locations
\cite{narducci2015,narducci}, rather than along distinct average momenta.

An interesting aspect that could be examined is the interplay between
which-path information and the sum over paths. While the probe shifts due to
the weak interactions can be understood as evidence that the quantum particle
went through both slits \cite{hofmann}, it is also possible to obtain
which-slit information by tagging into orthogonal polarisation states the
paths that came from distinct slits. Moreover, it should be possible to
implement a weak-value based partial quantum erasure procedure
\cite{mori2,caudano} to recover a partial interference.

To sum up, we have proposed a method based on minimally-perturbing weak
interactions in order to observe the paths superposition in a double-slit
interferometer. We have discussed an ideal scheme, based on inserting an array
of probes between the slits and the detection plane, as well as a simplified
protocol in view of an implementation of the method with single photons.

\section*{\uppercase{Appendix: measuring the single slit weak values in the photonic protocol}}%

\setcounter{equation}{0}

\renewcommand{\theequation}{A-\arabic{equation}} \noindent We consider 4
distinct experimental schemes in order to measure the weak values $\kappa
_{B}^{l}$ and $\kappa_{D}^{l}$ where $l=1,2$ designates the open slit. As in
the 2-slit setups, the initial polarization is taken to be $\left\vert
\nearrow\right\rangle $, the interaction Hamiltonians between the photon and
the crystals are $\hat{H}_{B}$ and $\hat{H}_{D}$ [see Eqs. (\ref{inistate})
and (\ref{hama})], and the post-selected also remains the same [Eq.
(\ref{defi99})]. Only the pre-selected state is of course different, given by
either $\psi_{p_{i}^{1}}^{1}(x,t=0)$ or $\psi_{p_{i}^{2}}^{2}(x,t=0)$.

With slit 1\ open and crystals B and D in place, in a first setup the
interaction seen by a photon is $\hat{H}_{B}+\hat{H}_{D}$; the resulting
polarization state after post-selection is%
\begin{equation}
|P_{1}^{\prime}\rangle=\langle\chi_{x_{f}}(t_{f})|\psi_{p_{i}^{1}}^{1}%
\rangle\big(e^{-i\gamma_{B}\kappa_{B}^{1}-i\gamma_{D}\kappa_{D}^{1}}%
|H\rangle+e^{i\gamma_{B}\kappa_{B}^{1}+i\gamma_{D}\kappa_{D}^{1}}%
|V\rangle\big).
\end{equation}
Then we change the setup by adding a phase-shifter to the D crystal and the
final polarization state is
\begin{equation}
|P_{2}^{\prime}\rangle=\langle\chi_{x_{f}}(t_{f})|\psi_{p_{i}^{1}}^{1}%
\rangle\big(e^{-i\gamma_{B}\kappa_{B}^{1}+i\gamma_{D}\kappa_{D}^{1}}%
|H\rangle+e^{i\gamma_{B}\kappa_{B}^{1}-i\gamma_{D}\kappa_{D}^{1}}%
|V\rangle\big).
\end{equation}
We repeat the same steps but with slit 2 open instead.\ The polarization
states are given by
\begin{align}
&  |P_{3}^{\prime}\rangle=\langle\chi_{x_{f}}(t_{f})|\psi_{p_{i}^{2}}%
^{2}\rangle\big(e^{-i\gamma_{B}\kappa_{B}^{2}-i\gamma_{D}\kappa_{D}^{2}%
}|H\rangle+e^{i\gamma_{B}\kappa_{B}^{2}+i\gamma_{D}\kappa_{D}^{2}}%
|V\rangle\big)\\
&  |P_{4}^{\prime}\rangle=\langle\chi_{x_{f}}(t_{f})|\psi_{p_{i}^{2}}%
^{2}\rangle\big(e^{-i\gamma_{B}\kappa_{B}^{2}+i\gamma_{D}\kappa_{D}^{2}%
}|H\rangle+e^{i\gamma_{B}\kappa_{B}^{2}-i\gamma_{D}\kappa_{D}^{2}}%
|V\rangle\big).
\end{align}
As in Sec. \ref{protox}, the polarization states are measured in the diagonal
basis.\ Defining the contrast $C_{i}^{\prime}$ similarly to $C_{i}$ in Eq.
(\ref{cont}), we find the polarization rotations are given by%
\begin{align}
&  \gamma_{B}\kappa_{1}^{B}=\frac{1}{2}\arccos(C_{1}^{\prime})+\frac{1}%
{2}\arccos(C_{2}^{\prime})\\
&  \gamma_{D}\kappa_{1}^{D}=\frac{1}{2}\arccos(C_{1}^{\prime})-\frac{1}%
{2}\arccos(C_{2}^{\prime})\\
&  \gamma_{B}\kappa_{2}^{B}=\frac{1}{2}\arccos(C_{3}^{\prime})+\frac{1}%
{2}\arccos(C_{4}^{\prime})\\
&  \gamma_{D}\kappa_{2}^{D}=\frac{1}{2}\arccos(C_{3}^{\prime})-\frac{1}%
{2}\arccos(C_{4}^{\prime}).
\end{align}

\end{document}